\documentclass[eqsecnum,noshowpacs,showkeys,nofootinbib,aps]{revtex4}
\usepackage{amsmath,amssymb}
\usepackage{epsfig}

\renewcommand{\theequation}{\arabic{equation}}
\newcommand\beq{\begin{equation}}
\newcommand\eeq{\end{equation}}
\newcommand\bea{\begin{eqnarray}}
\newcommand\eea{\end{eqnarray}}
\newcommand\nn{\nonumber}
\newcommand\pa{\partial}

\def\d{{\rm d}}

\renewcommand\d{{\rm d}}

\begin{document}
\title{Foliation, topology and nucleon charge profiles in hypersphere soliton model}
\author{Soon-Tae Hong}
\email{galaxy.mass@gmail.com}
\affiliation{Center for Quantum Spacetime and Department of Physics, Sogang University, Seoul 04107, Korea}
\date{\today}
\begin{abstract}
In the hypersphere soliton model (HSM), we study the geometrical inner structures and the ensuing charge distributions 
of the nucleons by exploiting the aspect of the HSM where the hypersphere soliton is 
described by an extended object possessing the parameter $\lambda$ $(0\le\lambda<\infty)$ which corresponds 
to the radial distance from the center of $S^{3}$ to the foliation leaves of the hypersphere soliton.
To do this, we investigate the foliation and topology related with geometry on a hypersphere 
described by $(\mu,\theta,\phi)$. Exploiting the so-called scanning algorithm we study geometrical relations 
between spherical shell foliation leave on a northern hemi-hypersphere $S^{3}_{+}$ and that on 
a flat equatorial solid sphere $E^{3}$ which contains the center of $S^{3}$. We then elucidate the physical 
meaning of $\mu$ in $S^{3}$ of radius $\lambda$ by showing that $\mu$ plays the role of an auxiliary angle 
to fix the radius $\lambda\sin\mu$ of the $S^{2}$ spherical shell sharing the center of 
$S^{3}(=S^{2}\times S^{1})$, at a given angle $\mu$. Next, using the charge density profiles of nucleons 
with $\mu$ dependence, we construct the nucleon fractional charges of spherically symmetric and nontrivial 
distributions. In the HSM we note that the proton and neutron charges do not leak out from the hypersphere soliton, 
and the positive and negative charges in the neutron are confined inside and outside its core, respectively. 
Explicitly we predict the fractional volumes and charges of the neutron. 
The proton and neutron are shown to be described by a topological structure of two 
Hopf-linked M\"obius strip type twist circles in $S^{3}$. We also note that the characteristic 
ratio of the hypersphere volume to the corresponding solid sphere one is given by a geometrical 
invariant related with hyper-compactness. 
\end{abstract}
\maketitle

\section{Introduction}
\setcounter{equation}{0}
\renewcommand{\theequation}{\arabic{section}.\arabic{equation}}

Exploiting the Dirac quantization in the first class formalism~\cite{di,hong15}, the hadron phenomenology theorists have tried 
to quantize constrained systems. In particular, the standard Skyrmion~\cite{skyrme61,anw83,hong98plb,hong15} 
has been studied as one of the constrained Hamiltonian systems. In this model, the three dimensional physical space is assumed 
to be topologically compactified to $S^{3}$ and the spatial infinity is located on the north-pole of $S^{3}$~\cite{anw83}. Next, 
the hypersphere soliton model (HSM)~\cite{manton1,hong98plb,hong21} has been proposed to find 
a set of equations of motion related with a topological lower bound on the 
soliton energy via the canonical quantization in the second class formalism~\cite{manton1}. 
Moreover, making use of the canonical quantization in the HSM, we have evaluated 
the baryon physical quantities such as baryon masses, charge radii and magnetic moments, 
to propose that a realistic hadron physics can be described by this phenomenological soliton 
on the hypersphere $S^{3}$~\cite{hong98plb}. Next, in the first class Hamiltonian quantization, we have evaluated the nucleon and delta baryon masses, 
charge radii, magnetic moments and axial coupling constant, most predicted values of which are in good 
agreement with the corresponding experimental data~\cite{hong21}.

Even though we have seen the phenomenological successes in predicting the baryon static 
properties~\cite{hong98plb,hong21} which are obtained after integrating out the angle 
$\mu$ of $S^{3}$ defined by the hypersphere coordinates $(\mu,\theta,\phi)$ in the HSM, 
it remains unclear to explain the exact physical meaning and role of $\mu$ associated with $S^{3}$ geometry. Note 
that some geometrical aspects of the static quantities of the baryons have been well explained~\cite{hong21} on 
the $S^{3}$ hypersphere in terms of the foliation~\cite{foli,hong21}. However the foliation related with the physical 
quantities such as the nucleon charge density profiles has not been studied in terms of $\mu$ of the 
three dimensional hypersphere $S^{3}$.

Next the internal structure of a nucleon is still a subject of great
interest to both experimentalists and theorists. In 1933, Frisch
and Stern~\cite{stern33} performed the first measurement of the
magnetic moment of proton, and obtained the earliest
experimental evidence for the internal structure of the nucleon. 
Since Coleman and Glashow~\cite{cg} predicted the baryon magnetic moments about sixty years ago, 
there has been a lot of progress in both theoretical paradigms and experimental
verifications for the magnetic moments. The measurement of
the baryon magnetic moment $\mu_{\Delta^{++}}^{\rm exp}=4.7-6.7$ has been reported~\cite{boss} 
to yield a new avenue for understanding hadron structure. Note that the theoretical prediction 
$\mu_{\Delta^{++}}=5.69$ in the HSM is within the corresponding experimental value~\cite{hong21}.

In this paper, we will investigate the geometrical and topological properties of the nucleon internal structure, 
by exploiting the HSM. 
In particular, using the characteristic of the HSM in which the hypersphere soliton is 
delineated by an extended object, we will scrutinize the geometrical inner structures and the ensuing charge distributions 
of the nucleons in terms of the parameter $\lambda$ $(0\le\lambda<\infty)$ which corresponds 
to the radial distance from the center of $S^{3}$ to the foliation leaves~\cite{hong21,foli} of the hypersphere soliton.
To accomplish this, we will consider the projection from 
a northern hemi-hypersphere $S^{3}_{+}$ to a three dimensional equatorial solid sphere $E^{3}$ which 
contains the center of $S^{3}$ and is described in terms of $(r, \theta, \phi)$.

From now on, we will name the mathematical algorithm to fill a compact manifold by using 
the foliation leaves on the manifold, scanning. Next, we will formulate the 
charge density profiles of nucleons which are given in terms of the angle $\mu$ in $S^{3}$. Making use of the 
geometrical descriptions on the three dimensional hypersphere soliton, we will find 
the nucleon charge distributions and the corresponding volumes which are spherically symmetric. We will also 
discuss two Hopf-linked~\cite{manton2,bott,kauffman,baez} M\"obius strip type circles in the HSM. Next we will briefly 
discuss the boson and fermion via the strip structures.

In Section 2, we will study the hypersphere geometry, foliation and scanning algorithm in the HSM. 
In Section 3, we will construct the three dimensional volumes associated with the spherically symmetric 
nucleon charge density profiles in $S^{3}$. In Section 4, in the HSM we will investigate the Hopf-links 
of the M\"obius strip type twist circles inside the baryons. Section 5 includes conclusions. 
In Appendix A, we will briefly study the formalism for the nucleon charges, 
mean square charge radii and magnetic moments.

\section{Foliation and scanning algorithm in $S^{3}$ geometry}
\setcounter{equation}{0}
\renewcommand{\theequation}{\arabic{section}.\arabic{equation}}

In this section, we will investigate the hypersphere $S^{3}$ geometry 
associated with the foliation~\cite{foli}, and the scanning algorithm in the HSM. 
To accomplish this, we briefly introduce the HSM~\cite{manton1,hong98plb,hong21} by starting with the Skyrmion Lagrangian density
\beq
{\cal L}=\frac{f_{\pi}^{2}}{4}{\rm tr}(\partial_{\mu}U^{\dagger}
         \partial^{\mu}U)+\frac{1}{32e^{2}}{\rm tr}[U^{\dagger}\partial_{\mu}U,
         U^{\dagger}\partial_{\nu}U]^{2},
\label{lagtot}
\eeq
where $U$ is an SU(2) chiral field, and $f_{\pi}$ and $e$ are a pion decay constant and a dimensionless Skyrme parameter, respectively. 
In this work, we will treat $f_{\pi}$ and $e$ as the model parameters~\cite{hong21,oka}. 
Next the quartic term is necessary to stabilize the soliton in the baryon sector.

Now we introduce the hypersphere manifold $S^{3}(=S^{2}\times S^{1})$. 
Note that $S^{3}$ itself is a three dimensional manifold. 
The three metric on $S^{3}$ described in terms of the hypersphere coordinates $(\mu,\theta,\phi)$ is given by
\beq
ds^{2}(S^{3})=\lambda^{2}d\mu^{2}+\lambda^{2}\sin^{2}\mu~(d\theta^{2}+\sin^{2}\theta~d\phi^{2}),
\label{metrics3}
\eeq
where the ranges of $(\mu,\theta,\phi)$  are given by $0\le\mu\le\pi$, $0\le\theta\le\pi$ and $0\le\phi\le 2\pi$, and 
$\lambda$ ($0\le \lambda <\infty$) is a radius of $S^{3}$. 
Note that the hypersphere $S^{3}$ is embedded in $R^{4}(=S^{3}\times R)$, 
and $R$ is a manifold associated with radial distance. We can then define the radius parameter $\lambda$ ($0\le \lambda <\infty$) 
as the radial distance from the center of $S^{3}$ to the foliation leaves~\cite{hong21, foli} of the hypersurface $S^{3}$ in $R^{4}$~\cite{hong21}.
Note also that from the three metric on $S^{3}$ in 
(\ref{metrics3}) we read off the radius of $S^{2}$ in $S^{3}(=S^{2}\times S^{1})$ which is 
given by $\lambda\sin\mu$ for a given angle $\mu$. This geometrical property of the radius $\lambda\sin\mu$ will be exploited 
later in the scanning algorithm in the hypersphere geometry.

On the $S^{3}$ hypersphere, the line element is defined as
\beq
d\vec{l}=\hat{e}_{\mu}\lambda~d\mu+\hat{e}_{\theta}\lambda\sin\mu~d\theta+\hat{e}_{\phi}\lambda\sin\mu\sin\theta~d\phi,
\label{dvecl}
\eeq
where $(\hat{e}_{\mu},\hat{e}_{\theta},\hat{e}_{\phi})$ are the unit vectors along the three directions. Exploiting 
the three arc length elements $\lambda d\mu$, $\lambda\sin\mu d\theta$ and $\lambda\sin\mu\sin\theta d\phi$ which 
can be read off from (\ref{dvecl}), we find the two dimensional area elements given by
\beq
d\vec{a}_{\mu}=\hat{e}_{\mu}\lambda^{2}\sin^{2}\mu\sin\theta~d\theta~d\phi,~~~
d\vec{a}_{\theta}=\hat{e}_{\theta}\lambda^{2}\sin\mu\sin\theta~d\mu~d\phi,~~~
d\vec{a}_{\phi}=\hat{e}_{\phi}\lambda^{2}\sin\mu~d\mu~d\theta,
\label{s3area}
\eeq
and the {\it three dimensional volume} element defined as
\beq
dV=\lambda^{3}\sin^{2}\mu\sin\theta~d\mu~d\theta~d\phi.
\label{volumnel}
\eeq

Making use of $d a_{\mu}$ in (\ref{s3area}), for a given angle $\mu$ we find the two dimensional area perpendicular 
to $\hat{e}_{\mu}$
\beq
A_{2}(S^{3})=4\pi\lambda^{2}\sin^{2}\mu.
\label{a2s3}
\eeq
Note that the surface area $A_{2}(S^{3})$ is a function of the angle $\mu$ and at $\mu=\frac{\pi}{2}$ it has a maximum value 
$A_{2}^{max}(S^{3})=4\pi\lambda^{2}$ which is equivalent to the surface area $A_{2}(S^{2})$ of the spherical shell $S^{2}$ of 
radius $\lambda$, as expected. Next exploiting $dV$ in (\ref{volumnel}), for the range $0\le\mu\le\mu_{0}$ we have the three dimensional volume 
\beq
V(S^{3}; 0\le\mu\le\mu_{0})=2\pi\lambda^{3}\left(\mu_{0}-\frac{1}{2}\sin 2\mu_{0}\right).
\label{a3s32}
\eeq
In particular, inserting $\mu_{0}=\pi$ into (\ref{a3s32}), we obtain the three dimensional volume of 
the hypersphere for the range $0\le\mu\le\pi$ 
\beq
V(S^{3})=2\pi^{2}\lambda^{3}.
\label{a3s3}
\eeq

Now we have comments on the three dimensional volume of 
the hypersphere $V(S^{3})$ in (\ref{a3s3}). We obtain 
$V(S^{3})=2\pi^{2}\lambda^{3}$ in (\ref{a3s3}) by using the two dimensional area $A_{2}(S^{3})$ in 
(\ref{a2s3}) and the line element $\lambda d\mu$
\beq
V(S^{3})=\int_{0}^{\pi} A_{2}(S^{3})\lambda d\mu.
\label{a3s3dmu}
\eeq
Moreover the three dimensional volume $V(S^{3})$ of the hypersphere of radius $\lambda$ is much larger than 
the three dimensional volume $V(R^{3})=\frac{4\pi}{3}\lambda^{3}$ of the solid sphere of 
radius $\lambda$ to yield the characteristic ratio $\chi$ of the form
\beq
\chi=\frac{V(S^{3})}{V(R^{3})}=\frac{3\pi}{2},
\label{chiratio}
\eeq
which is dimensionless and greater than unity.
Note that the hypersphere of radius $\lambda$ is delineated by the compact manifold which corresponds 
to the three dimensional volume $V(S^{3})$ in (\ref{a3s3}) and the ratio $\chi$ in (\ref{chiratio}).

\begin{figure}[t]
\centering
\vskip -0.5cm 
\includegraphics[width=7.0cm]{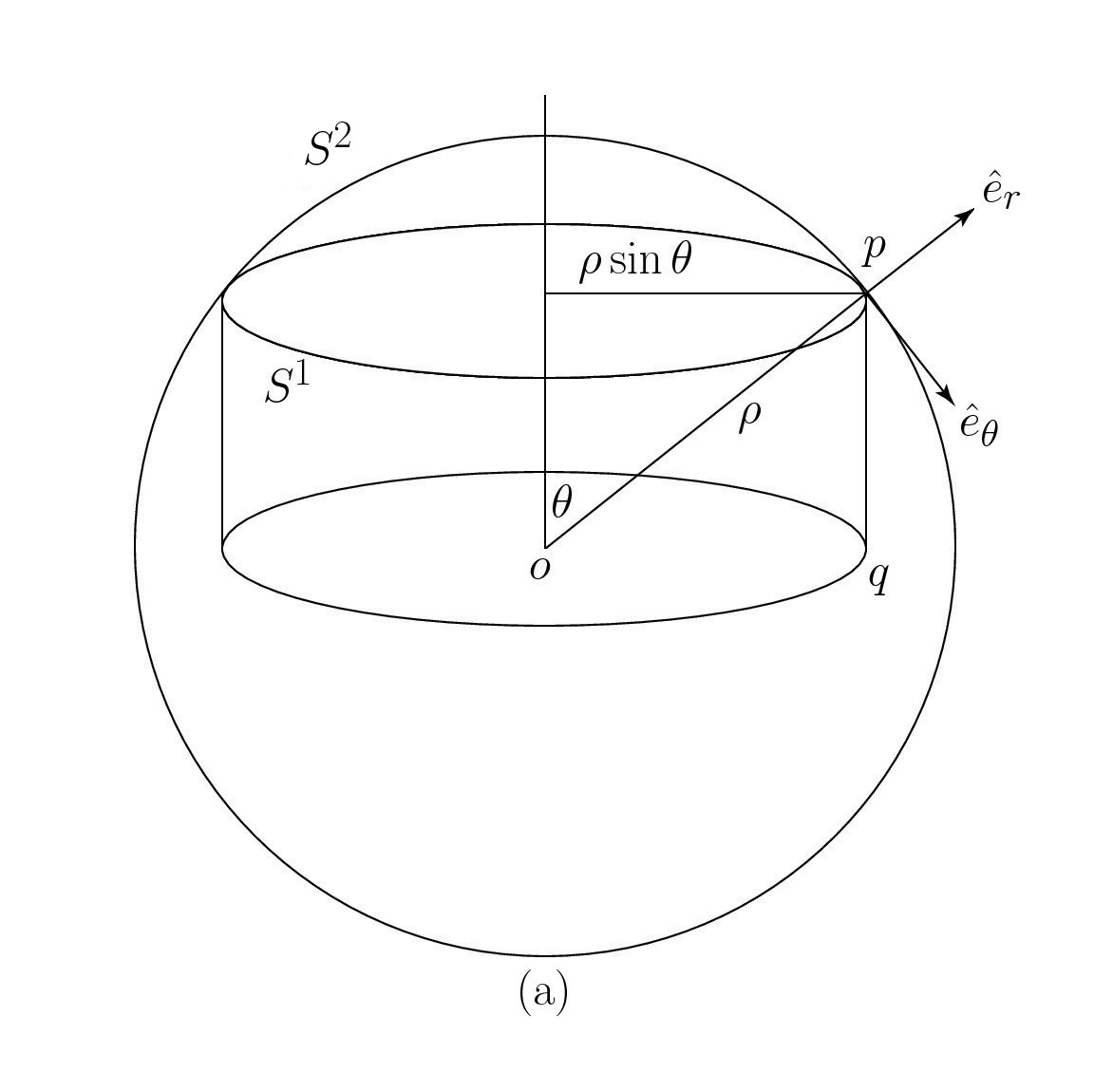}
\hskip 1.0cm
\includegraphics[width=7.0cm]{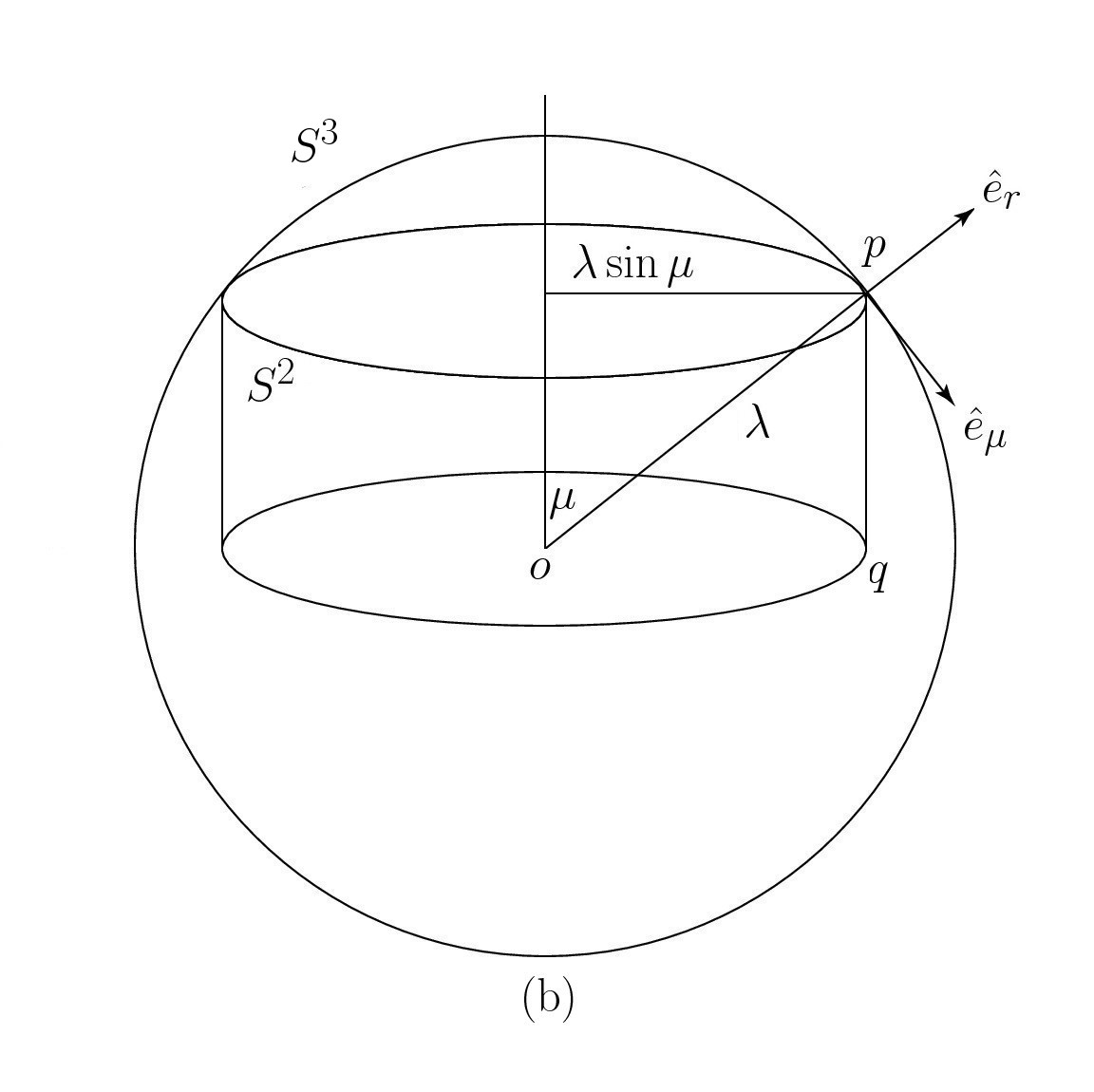}
\vskip -0.5cm 
\caption[circle] {The schematic scanning algorithm diagrams for (a) the two dimensional spherical shell $S^{2}$ geometry and (b) 
the three dimensional hypersphere $S^{3}$ one depicted in terms of $S^{2}\times S^{1}$, 
where $S^{2}$ is simplified by $S^{1}$ for convenience. In (b) the perpendicular unit vectors 
along the three directions are described by $(\hat{e}_{\mu}, \hat{e}_{\theta}, \hat{e}_{\phi})$, where 
the unit vectors $\hat{e}_{\theta}$ and $\hat{e}_{\phi}$ are perpendicular to each other even though 
they cannot be explicitly depicted in this schematic diagram. Note that 
the unit vector $\hat{e}_{r}$ is perpendicular to the spherical shell $S^{2}$ which is described by 
$(\hat{e}_{\theta}, \hat{e}_{\phi})$ and shares the center $o$ of $S^{3}$. For more details of the 
spherical shell $S^{2}$ geometry in (b), see the text in Section 2.} 
\label{circle}
\end{figure}

Next we investigate the foliation related with the geometry inside the hypersphere. 
To do this, we first consider a toy model of the spherical shell $S^{2}$ of radius $\rho$ which is described in terms of $(\theta, \phi)$, 
as shown in Fig.~\ref{circle}(a). 
Here the perpendicular unit vectors along the two directions are denoted by 
$(\hat{e}_{\theta},\hat{e}_{\phi})$.
Note that $\hat{e}_{\phi}$ 
is into the page at the observation point $p$. On the {\it curved} northern hemisphere $S^{2}_{+}$, 
we have foliation leaves of circle of radius $\rho\sin\theta$ $(0\le\theta\le\frac{\pi}{2})$. Note that 
these foliation leaves fill $S^{2}_{+}$ after scanning completely over the range $0\le\theta\le\frac{\pi}{2}$. 
Similarly the southern hemisphere $S^{2}_{-}$ is filled with the foliation leaves of circle of radius 
$\rho\sin\theta$ $(\frac{\pi}{2}\le\theta\le\pi)$. Completing the
scanning over the range 
$0\le\theta\le\pi$ we find the two dimensional surface area $A_{2}(S^{2})=4\pi\rho^{2}$. 
We next consider a projection from the northern hemisphere $S^{2}_{+}$ 
to the two dimensional equatorial plane $E^{2}$ which contains the center of $S^{2}$ 
as shown in Fig.~\ref{circle}(a). To be specific, the circle 
of radius $\rho\sin\theta$ on the curved manifold $S^{2}_{+}$ is projected on the circle of radius $\rho\sin\theta$ 
on the equatorial plane $E^{2}$. In other words, the radius $\rho\sin\theta$ on $S^{2}_{+}$ is projected on the radial
distance $oq$ on $E^{2}$. 
As we exploit the foliation leaves of circle of radius $\rho\sin\theta$ to scan the curved manifold $S^{2}_{+}$ over the 
range $0\le\theta\le\frac{\pi}{2}$, we also scan 
the flat equatorial plane $E^{2}$ with the foliation leaves of circle of radius $\rho\sin\theta$ to produce the surface area 
$A_{2}(E^{2})=\pi \rho^{2}$ which is less than $A_{2}(S^{2}_{+})=2\pi \rho^{2}$. Note that 
$A_{2}(S^{2})=A_{2}(S^{2}_{+})+A_{2}(S^{2}_{-})=4\pi \rho^{2}$.

Now we apply the same mathematical logic used in the scanning algorithm for the toy model $S^{2}$, 
to the three dimensional volume of the hypersphere $S^{3}$ of radius $\lambda$ which is described 
in terms of $(\mu, \theta, \phi)$. To do this, we exploit the 
hypersphere $S^{3}$ geometry schematically depicted in terms of $S^{2}\times S^{1}$ as shown in Fig.~\ref{circle}(b), where 
$S^{2}$ is simplified by $S^{1}$ for convenience. 
Here the perpendicular unit vectors along the three directions are denoted by 
$(\hat{e}_{\mu}, \hat{e}_{\theta}, \hat{e}_{\phi})$.
Moreover, the unit vectors $\hat{e}_{\theta}$ and $\hat{e}_{\phi}$ are perpendicular to each other even though they cannot be explicitly 
depicted in the {\it schematic diagram} in Fig.~\ref{circle}(b).
Note that, because $S^{3}$ is embedded in $R^{4}(=S^{3}\times R)$, $S^{3}$ is delineated in terms of the 
unit vectors $(\hat{e}_{\mu}, \hat{e}_{\theta}, \hat{e}_{\phi})$ and $R$ is described by the unit vector $\hat{e}_{r}$. 
Moreover the radius of the hypersphere is given by $\lambda$, which is {\it a radius parameter of the manifold $R$.}
\footnote{Since $S^{2}$ is embedded in $R^{3}(=S^{2}\times R)$, $S^{2}$ is described in terms of the unit 
vectors $(\hat{e}_{\theta}, \hat{e}_{\phi})$ and $R$ is delineated by the unit vector $\hat{e}_{r}$. 
Furthermore the radius of the spherical shell is given by $\rho$, namely a radius parameter of the manifold $R$. 
The same logic can be applied to $S^{3}$ of radius parameter $\lambda$ embedded in $R^{4}(=S^{3}\times R)$.}

Next we consider a projection from the northern hemi-hypersphere $S^{3}_{+}$ 
to the equatorial solid sphere 
$E^{3}$, which contains the center of $S^{3}$ and is described by the schematic planar disk in Fig.~\ref{circle}(b). 
At a given angle $\mu$, the spherical shell of radius $\lambda\sin\mu$ on the curved manifold $S^{3}_{+}$ 
is projected on the spherical shell of radius $\lambda\sin\mu$ on the flat equatorial solid sphere $E^{3}$. 
In other words, the radius $\lambda\sin\mu$ on $S^{3}_{+}$ is projected on the radial distance $oq$ on $E^{3}$ in Fig.~\ref{circle}(b).

Now we have comments on the physical meaning and role of $\mu$ in the hypersphere $S^{3}$. 
Exploiting (\ref{metrics3}) at a {\it fixed angle} $\mu$ in $S^{3}$, we have the two metric 
$ds^{2}=\lambda^{2}\sin^{2}\mu~(d\theta^{2}+\sin^{2}\theta~d\phi^{2})$ so that we can obtain a concentric $S^{2}$ 
spherical shell (or foliation leave) of radius $\lambda\sin\mu$, sharing the center $o$ of the hypersphere 
$S^{3}(=S^{2}\times S^{1})$ in Fig.~\ref{circle}(b).\footnote{In the $S^{2}$ toy model, the 
$S^{1}$ circles (or foliation leaves) of radius $\rho\sin\theta$ {\it do not share} the center $o$ of the sherical shell $S^{2}$ 
as in Fig.~\ref{circle}(a).} Here the radius of the foliation leave 
$S^{2}$ at the fixed angle $\mu$ is independent of the angles $(\theta, \phi)$ to produce the radial distance 
$\lambda\sin\mu$. This discussion on the two metric at a fixed angle 
$\mu$ in $S^{3}$ is consistent with the scanning algorithm for the $S^{3}$ geometry 
which is schematically depicted in Fig.~\ref{circle}(b). As a result, $\mu$ can be interpreted as an auxiliary angle 
to assign to a given angle $\mu$ of $S^{1}$, the radius $\lambda\sin\mu$ of the $S^{2}$ spherical shell sharing the center of 
$S^{3}$. Note that, as shown in (\ref{dvecl}), $\hat{e}_{\mu}$ of the line element 
$\lambda d\mu$ in $S^{1}$ is perpendicular to ($\hat{e}_{\theta}, \hat{e}_{\phi}$) of the two dimensional area 
$A_{2}(S^{3})$ of radius $\lambda\sin\mu$. Note also that, since the integrand $A_{2}(S^{3})$ ($0\le\mu\le\pi$) of 
$V(S^{3})$ in (\ref{a3s3dmu}) consists of the foliation leaves sharing the center of $S^{3}$, 
$\hat{e}_{r}$ in Fig.~\ref{circle}(b) is perpendicular to $S^{2}$ whose center is located at the origin $o$ 
of $S^{3}$.

On the northern hemi-hypersphere $S^{3}_{+}$, for a fixed $\mu$ we construct a foliation leave $S^{2}$ of 
radius $\lambda\sin\mu$ having an area 
$A_{2}(S^{3})=4\pi\lambda^{2}\sin^{2}\mu$ in (\ref{a2s3}). 
Completing the scanning over the range $0\le\mu\le\frac{\pi}{2}$ 
corresponding to $S^{3}_{+}$ and using (\ref{a3s32}), we find the three dimensional volume 
$V(S_{+}^{3})=\pi^{2}\lambda^{3}$. Similarly, we scan over the range $\frac{\pi}{2}\le\mu\le\pi$ related with $S^{3}_{-}$ 
to yield the volume $V(S_{-}^{3})=\pi^{2}\lambda^{3}$. As a result, 
we scan over the range $0\le\mu\le\pi$ associated with the total manifold $S^{3}$ to yield 
\beq
V(S^{3})=V(S^{3}_{+})+V(S^{3}_{-})=2\pi^{2}\lambda^{3},
\label{a3s3sum}
\eeq 
which produces the geometrical ratio $\chi$ in (\ref{chiratio}), and 
is consistent with (\ref{a3s3}) obtained by exploiting the pure geometrical approach 
without resorting to the scanning algorithm. Note that this consistency of the volumes in 
(\ref{a3s3}) and (\ref{a3s3sum}) confirms the above statements on the role of $\mu$ as an auxiliary angle.

Next the foliation leave of radius $\lambda\sin\mu$ on $S^{3}_{+}$ 
is distinct from that on the equatorial solid sphere $E^{3}$, since the former is attached to the {\it curved} manifold 
$S^{3}_{+}$ while the latter adheres to the {\it flat} one $E^{3}$. 
Exploiting the projection from $S^{3}_{+}$ with the spherical shells of 
radius $\lambda\sin\mu$ $(0\le\mu\le\frac{\pi}{2})$ to $E^{3}$ with 
those of radius $\lambda\sin\mu$, we find the volume $V(R^{3})=\frac{4\pi}{3}\lambda^{3}$ 
which is less than $V(S^{3}_{+})=\pi^{2}\lambda^{3}$. Note that there exists a difference between the 
scanning on $S^{3}$ and that in $E^{3}$. To be specific, on $S^{3}$ we scan {\it both} over $S^{3}_{+}$ and over $S^{3}_{-}$, while 
in $E^{3}$ we scan {\it only once} over the schematic planar disk in Fig.~\ref{circle}(b). 
The characteristic invariant $\chi$ in (\ref{chiratio}) thus implies that the volume $V(S^{3})$ 
in (\ref{a3s3sum}) is {\it hyper-compact} with the ratio $\chi=\frac{3\pi}{2}~(>1)$ with respect to 
the corresponding solid sphere volume $V(R^{3})$. In other words, in the HSM the baryon in the strong interaction is described in 
terms of the geometrical binding related with the hyper-compact manifold volume $V(S^{3};\lambda=\lambda_{B})$ 
inside which the baryon charges are confined.

\section{Nucleon charge distribution profiles in HSM}
\setcounter{equation}{0}
\renewcommand{\theequation}{\arabic{section}.\arabic{equation}}

In this section, we will investigate the charge profile of nucleon of a fixed radius of $\lambda_{B}$.
To explicitly find $\lambda_{B}$, we study the charge radii. In the HSM, 
we construct the magnetic isovector mean square charge radius~\cite{hong98plb,hong21}
\begin{equation}
\langle r^{2}\rangle_{M,I=1}=\frac{2}{3e^{2}{\cal I}}\int_{S^{3}}dV_{B}\sin^{2}\mu\sin^{2}f
\left(1+\left(\frac{d f}{d\mu}\right)^{2}+\frac{\sin^{2}f}{\sin^{2}\mu}\right)=\frac{5}{6e^{2}f_{\pi}^{2}},
\label{r2exp21}
\end{equation}
where the subscript $M$ denotes magnetic charge radius, ${\cal I}$ is the moment of inertia, and $dV_{B}=\lambda_{B}^{3}\sin^{2}\mu\sin\theta~d\mu~d\theta~d\phi$ 
on the hypersphere $S^{3}$. Note that $dV_{B}$ is given by product of three arc lengths: $\lambda_{B}d\mu$, $\lambda_{B}\sin\mu d\theta$ and 
$\lambda_{B}\sin\mu\sin\theta d\phi$ and $\lambda_{B}$ is radius of hypersphere soliton. Moreover, we find the charge radii in 
terms of $ef_{\pi}$~\cite{hong98plb,hong21}
\begin{eqnarray}
\langle r^{2}\rangle^{1/2}_{M,I=0}&=&\langle r^{2}\rangle^{1/2}_{M,I=1}
=\langle r^{2}\rangle^{1/2}_{M,p}=\langle r^{2}\rangle^{1/2}_{M,n}=\langle r^{2}\rangle^{1/2}_{E,I=1}
=\sqrt{\frac{5}{6}}\frac{1}{ef_{\pi}},\nonumber\\
\langle r^{2}\rangle^{1/2}_{E,I=0}&=&\frac{\sqrt{3}}{2}\frac{1}{ef_{\pi}},~~~
\langle r^{2}\rangle_{p}=\frac{19}{24}\frac{1}{(ef_{\pi})^{2}},~~~
\langle r^{2}\rangle_{n}=-\frac{1}{24}\frac{1}{(ef_{\pi})^{2}},
\label{radii4}
\end{eqnarray}
where the subscript $E$ stands for electric charge radii. For more details of the formalism for the mean square charge radii and magnetic moments, see Appendix A.

\begin{table}[t]
\caption{In Prediction~\cite{hong21}, we use the hypersphere soliton model in the first class Dirac formalism with the WOC. 
The input parameters are indicated by $*$.}
\begin{center}
\begin{tabular}{lrrrrr}
\hline
Quantity   &~~~~~~Prediction &~~~~~~Experiment &~~~~~~Quantity   &~~~~~~Prediction &~~~~~~Experiment\\
\hline
$\langle r^{2}\rangle^{1/2}_{M,I=0}$ &0.80 {\rm fm} &0.81 {\rm fm} &$\langle r^{2}\rangle^{1/2}_{E,I=0}$  &0.76 {\rm fm} &0.72 {\rm fm}\\ 
$\langle r^{2}\rangle^{1/2}_{M,I=1}$ &0.80 {\rm fm} &0.80 {\rm fm} &$\langle r^{2}\rangle^{1/2}_{E,I=1}$  &0.80 {\rm fm} &0.88 {\rm fm}\\ 
$\langle r^{2}\rangle^{1/2}_{M,p}$  &0.80 {\rm fm$^{*}$} &0.80 {\rm fm} &$\langle r^{2}\rangle_{p}$  &(0.780 {\rm fm})$^{2}$ &(0.805 {\rm fm})$^{2}$\\
$\langle r^{2}\rangle^{1/2}_{M,n}$  &0.80 {\rm fm}  &0.79 {\rm fm} &$\langle r^{2}\rangle_{n}$  &$-(0.179~{\rm fm})^{2}$ &$-(0.341~{\rm fm})^{2}$\\
$\mu_{p}$  &2.98 &2.79 &$M_{N}$ &939 {\rm MeV$^{*}$} &939 {\rm MeV}\\
$\mu_{n}$  &$-2.45$ &$-1.91$ &$M_{\Delta}$ &1112 {\rm MeV} &1232 {\rm MeV}\\
$\mu_{\Delta^{++}}$  &5.69  &$4.7-6.7$ &$g_{A}$ &1.30 &$1.23$\\
\hline
\end{tabular}
\end{center}
\label{tablestatic}
\end{table}

Inserting the experimental value $\langle r^{2}\rangle^{1/2,{\rm exp}}_{M,I=1}=0.80~{\rm fm}$~\cite{liu87,hong98plb,hong21} 
into (\ref{radii4}), we find $ef_{\pi}=(0.876~{\rm fm})^{-1}$ which corresponds to 
\beq
\lambda_{B}=0.876~{\rm fm}.
\label{lambdab}
\eeq 
In other words, the hypersphere soliton is defined in terms of  the fixed radius $\lambda_{B}=0.876~{\rm fm}$ in (\ref{lambdab}) 
which is comparable to the charge radius $\langle r^{2}\rangle^{1/2}_{M,I=1}=0.80~{\rm fm}$ predicted in the HSM~\cite{hong21}. 
Note that the physical quantity $\langle r^{2}\rangle^{1/2}_{M,I=1}$ in (\ref{radii4}), which is obtained after 
integrating out the angle $\mu$ in (\ref{r2exp21}) and is an invariant independent of $\mu$, can be 
phenomenologically compared with the corresponding experimental value related with the 
charge effect of the hypersphere soliton.

Next we discuss the estimation of the HSM. To do this, we exploit the above value of $ef_{\pi}=(0.876~{\rm fm})^{-1}$ 
and the charge radii in (\ref{radii4}). The predictions for the physical quantities such as the charge radii in (\ref{radii4}), 
magnetic moments, baryon masses and axial coupling constant 
are listed in Table~I~\cite{hong21}. In particular, we exploit the first class Dirac quantization~\cite{hong21} and Weyl 
ordering correction (WOC) formalism~\cite{hong21,lee} to obtain $M_{N}$ and $M_{\Delta}$ 
in (\ref{mnmdelta}). Note that the predicted values for 
$\mu_{\Delta^{++}}$, $\langle r^{2}\rangle^{1/2}_{M,I=0}$, 
$\langle r^{2}\rangle^{1/2}_{M,I=1}$, $\langle r^{2}\rangle^{1/2}_{M,n}$ as well as the input parameters $M_{N}$ and 
$\langle r^{2}\rangle^{1/2}_{M,p}$ are almost the same as the corresponding experimental data. Next the predictions 
for $g_{A}$, $\langle r^{2}\rangle^{1/2}_{E,I=0}$ and $\langle r^{2}\rangle_{p}$ 
($M_{\Delta}$, $\mu_{p}$ and $\langle r^{2}\rangle^{1/2}_{E,I=1}$)   
are within about 6 \% (10 \%) of the experimental values. 
In contrast, the estimation of the standard Skyrmion model has been known to be not in good agreement with the corresponding 
experimental data~\cite{anw83,liu87}. In particular, in this model the predictions
of the charge radii are as follows: $\langle r^{2}\rangle^{1/2}_{M,I=1}$=$\langle r^{2}\rangle^{1/2}_{M,p}$=
$\langle r^{2}\rangle^{1/2}_{M,n}$=$\langle r^{2}\rangle^{1/2}_{E,I=1}$=$\langle r^{2}\rangle_{p}$=$\infty$ and 
$\langle r^{2}\rangle_{n}$=$-\infty$~\cite{anw83,liu87}.

Now we investigate the physical meaning and role of $\mu$ on the $S^{3}$ geometry of the nucleon in the phenomenological HSM. 
To accomplish this, we first consider the hypersphere of radius $\lambda_{B}$ having the 
three dimensional volume $V(S^{3};\lambda=\lambda_{B})$ which is given with the 
specific value $\lambda=\lambda_{B}$ in (\ref{a3s3}). 
Explicitly, exploiting $\lambda_{B}=0.876~{\rm fm}$ in (\ref{lambdab}), 
we find the compact manifold volume $V(S^{3};\lambda=\lambda_{B})$ possessing the characteristic ratio 
$\chi=\frac{3\pi}{2}$ in (\ref{chiratio})
\beq
V(S^{3};\lambda=\lambda_{B})=2\pi^{2}\lambda_{B}^{3}=13.269~{\rm fm}^{3},
\label{volumeblambda}
\eeq
which is greater than the corresponding solid sphere volume $V(R^{3};\lambda=\lambda_{B})=2.816~{\rm fm}^{3}$. 
Note that the universe ${\cal U}$ in the HSM is given by a sum of two subspaces, namely 
${\cal U}=V(S^{3};\lambda=\lambda_{B})\oplus\bar{V}(S^{3};\lambda=\lambda_{B})$ where $V(S^{3};\lambda=\lambda_{B})$ and 
$\bar{V}(S^{3};\lambda=\lambda_{B})$ are the volume subspaces inside and outside the hypersphere soliton, respectively.

\begin{figure}[t]
\centering
\includegraphics[width=7.0cm]{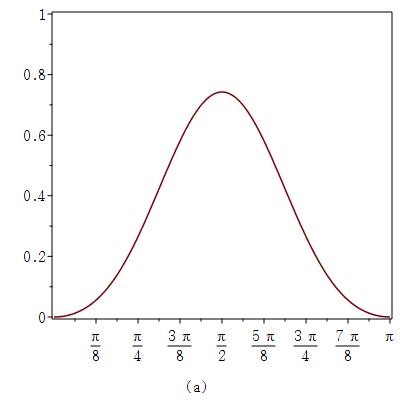}
\hskip 0.2cm
\includegraphics[width=7.0cm]{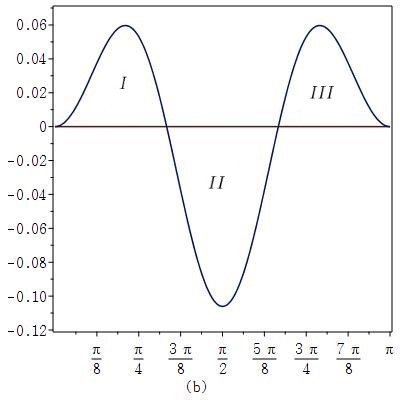}
\caption[chargep] {The charge density profiles 
(a) $\rho_{p}(\mu)$ and (b) $\rho_{n}(\mu)$, for proton and neutron respectively, are plotted versus $\mu$. 
Here $\rho_{p}(\mu)$ and $\rho_{n}(\mu)$ are associated with the charges inside the nucleons.} 
\label{chargep}
\end{figure}

Next on the hypersphere we obtain the proton and neutron charges as follows
\beq
Q_{p,n}=\frac{e^{3}f_{\pi}^{3}}{4\pi^{2}}\int_{S^{3}}dV_{B}\left(1\pm \frac{4}{3}\sin^{2}\mu\right).
\label{pncharge}
\eeq
Note that, after performing the integration of the nucleon charges $Q_{p,n}$ over the hypersphere $S^{3}$, 
we find the desired results $Q_{p}=1$ and $Q_{n}=0$. Note also that the integrands in (\ref{r2exp21}) and (\ref{pncharge}) 
do not possess any dependences on the coordinates $(\theta,\phi)$, and thus they are spherically symmetric physical 
densities on the hypersphere. 
To be specific, the integrands in (\ref{r2exp21}) and (\ref{pncharge}) have a 
dependence on $\mu$, which is one of the parameters of $S^{3}$, and thus they are not hyper-spherically symmetric. 
Moreover, after integrating these integrands over the solid angle associated 
with $(\theta,\phi)$, we find the corresponding spherically symmetric physical quantities.
The proton and neutron charge 
densities corresponding to the integrand of the integral in (\ref{pncharge}) thus have the spherical 
symmetries in the HSM. Integrating the integrand of $Q_{p,n}$ in (\ref{pncharge}) over the solid angle, we arrive at
\beq
Q_{p,n}=\int_{0}^{\pi}d\mu\rho_{p,n}(\mu),
\label{pncharge2}
\eeq 
where the charge density profiles $\rho_{p}(\mu)$ and $\rho_{n}(\mu)$ are given by
\bea
\rho_{p}(\mu)&=&\frac{1}{\pi}\sin^2\mu \left(1+\frac{4}{3}\sin^2 \mu\right),\nn\\
\rho_{n}(\mu)&=&\frac{1}{\pi}\sin^2\mu \left(1-\frac{4}{3}\sin^2 \mu\right).
\label{rhopn}
\eea
For more details of the charge operator and charge density profiles, see Appendix A.
Note that $\rho_{p}(\mu)$ and $\rho_{n}(\mu)$ are spherically symmetric. The proton and neutron charge densities in  (\ref{rhopn}) 
are then effectively depicted in terms of the angle $\mu$ only as shown in Fig.~\ref{chargep}(a) and Fig.~\ref{chargep}(b), respectively.

Now we have some comments on the nucleon charge distributions on $S^{3}$ described by $\mu$. 
First, since $A_{2}(S^{3};\lambda=\lambda_{B})$ in (\ref{a3s3dmu}) is perpendicular to $\lambda_{B}d\mu$ 
as shown in (\ref{dvecl}), the arc length element $\lambda_{B} d\mu$ in the spherically symmetric hypersphere $S^{3}$ of radius 
$\lambda_{B}$ plays a role similar to the line element $dr$ in the spherically symmetric equatorial solid sphere $E^{3}$. 
Since we have the three dimensional volume on $S^{3}$ soliton, we thus treat 
$dV_{B}=\lambda_{B}^{3}\sin^{2}\mu\sin\theta~d\mu~d\theta~d\phi$ as a volume element inside the hypersphere 
describing the nucleon, implying that the $\mu$ dependences in $\rho_{p}(\mu)$ and $\rho_{n}(\mu)$ are related with 
the radial dependences in the charge densities inside the nucleons. 

Second, we define $\lambda_{B}$ as the radial distance from the center of $S^{3}$ to the boundary of the 
hypersphere $S^{3}$~\cite{hong21}. Note that the calculated charge radius $\langle r^{2}\rangle^{1/2}_{M,I=1}$ in (\ref{radii4}) 
can be defined as the fixed radial distance to the point on a manifold which does not need to be located only 
on the compact manifold $S^{3}$ of radius $\lambda_{B}$. To be specific, the magnetic isovector charge radius 
$\langle r^{2}\rangle_{M,I=1}^{1/2}$ denotes the radial distance which is a geometrical 
invariant~\cite{hong21}. 

Third, we can also define the radial distance inside the hypersphere, and this distance is related with the charge density inside 
the nucleon. Next the physical quantities $Q_{p}$ and $Q_{n}$ can be compared with the corresponding 
experimental values $Q_{p}^{\rm exp}=1$ and $Q_{n}^{\rm exp}=0$, similar to the other physical quantities 
such as the nucleon and delta baryon masses $M_{N}$ and $M_{\Delta}$
whose predictions and experimental data are given in Table~I. 

\begin{figure}[t]
\centering
\vskip -0.5cm
\hskip 0.5cm
\includegraphics[width=6.1cm]{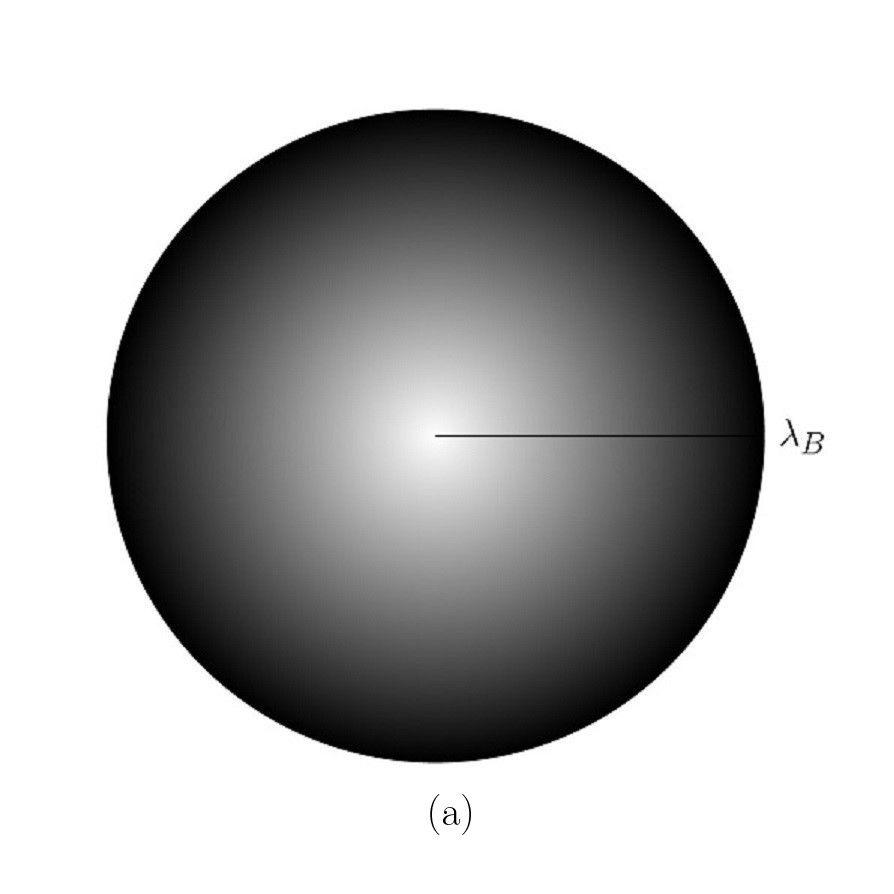}
\hskip 1.0cm
\includegraphics[width=6.1cm]{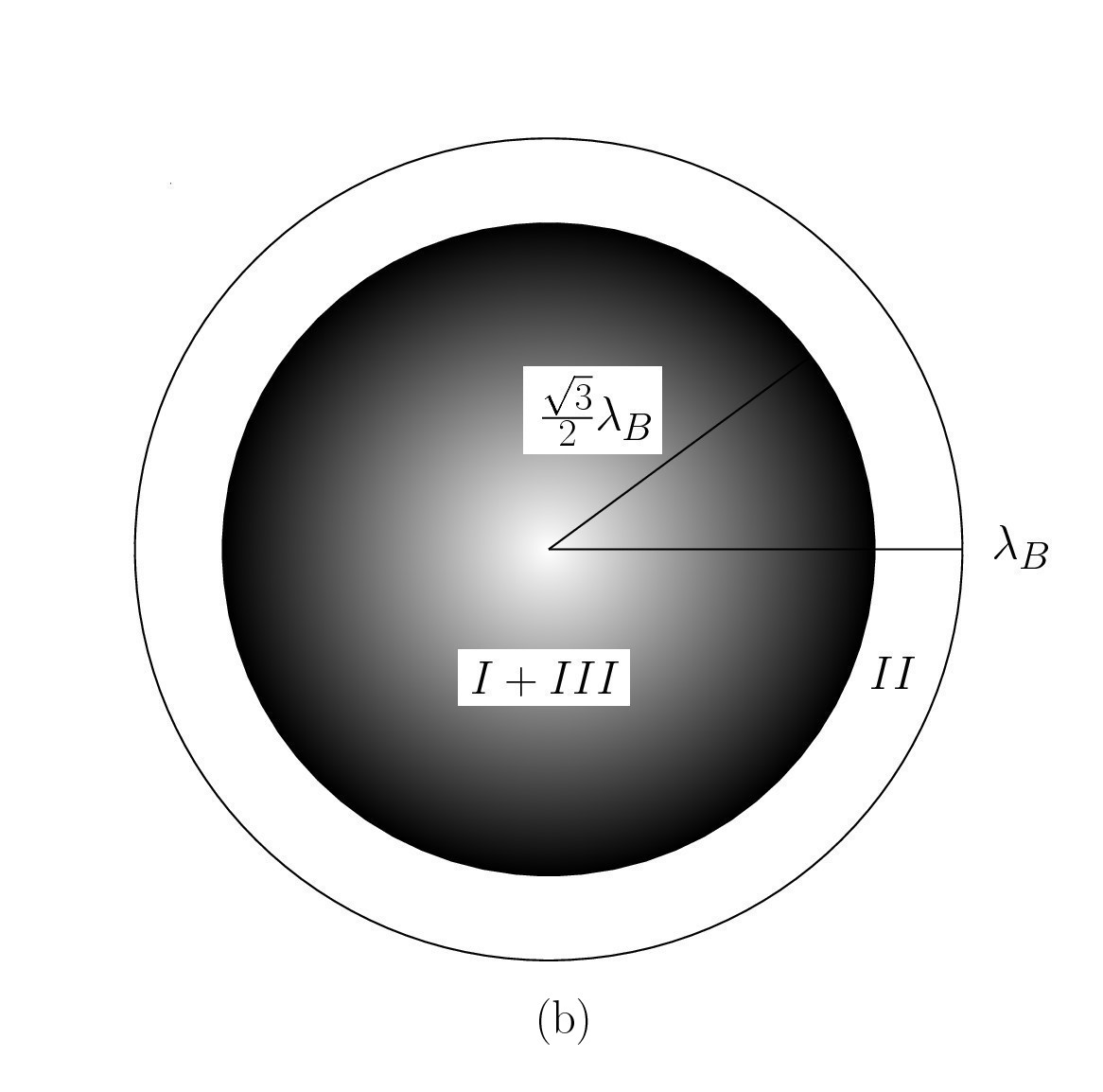}
\vskip -0.5cm 
\caption[hyper] {The schematic charge distributions for (a) the proton and (b) the neutron in the hypersphere 
of radius $\lambda_{B}$. In (a) the positive charge $Q_{p}=1$ fills the proton. In (b) the positive charge
$Q_{n}^{I}+Q_{n}^{III}=\frac{\sqrt{3}}{8\pi}$ is confined 
inside the core of the neutron, while the negative charge $Q_{n}^{II}=-\frac{\sqrt{3}}{8\pi}$ 
is distributed outside the core. Here these nucleon charges do not leak out from the hypersphere soliton.} 
\label{hyper}
\end{figure}

Next, exploiting the neutron charge density in (\ref{rhopn}), we find two root values of 
$\rho_{n}(\mu)$=0: $\mu=\frac{\pi}{3}$ and $\mu=\frac{2\pi}{3}$. 
We now have three regions $I$, $II$ and $III$ in the neutron charge density as shown in Fig.~\ref{chargep}(b), and for each region 
we obtain the charge fractions as follows
\beq
Q_{n}^{I}=Q_{n}^{III}=\frac{\sqrt{3}}{16\pi},~~~
Q_{n}^{II}=-\frac{\sqrt{3}}{8\pi},
\label{qn123}
\eeq
which are in good agreement with the result $Q_{n}=0$. Note that the charge density profiles of proton and neutron in (\ref{rhopn}) 
have no dependence on the coordinates $\theta$ and $\phi$ so that they can be spherically symmetric. Moreover 
all the charges fill the three dimensional volume $V(S^{3};\lambda=\lambda_{B})=2\pi^{2}\lambda_{B}^{3}$ of 
the hypersphere of radius $\lambda_{B}$. The proton charge is confined inside 
the hypersphere of radius $\lambda_{B}$ with positive charge only as shown in Fig.~\ref{hyper}(a). In contrast, the positive charge 
$Q_{n}^{I}+Q_{n}^{III}$ of the neutron is located inside the core of the hypersphere of radius $\frac{\sqrt{3}}{2}\lambda_{B}$ 
corresponding to the boundaries at $\mu=\frac{\pi}{3}$ and $\mu=\frac{2\pi}{3}$, while the neutron negative charge $Q_{n}^{II}$ 
is distributed outside the core, as shown in Fig.~\ref{chargep}(b) and Fig.~\ref{hyper}(b). 
Note that the proton and neutron charges do not leak out from the soliton, and similar to $\lambda_{B}$ the radial distance 
$\frac{\sqrt{3}}{2}\lambda_{B}$ in Fig.~\ref{hyper}(b) is a geometrical invariant.

Exploiting the geometrical descriptions on the hypersphere soliton, we evaluate the three dimensional volume associated with 
the neutron charge distributions possessing the three pieces in (\ref{qn123}) which are spherically symmetric. To be specific, making use of 
(\ref{a3s32}), we arrive at the three dimensional volumes for the fractional neutron distributions
\bea
V_{n}^{I}(S^{3};\lambda=\lambda_{B})&=&V_{n}^{III}(S^{3};\lambda=\lambda_{B})=2\pi\lambda_{B}^{3}\left(\frac{\pi}{3}-\frac{\sqrt{3}}{4}\right),\nn\\
V_{n}^{II}(S^{3};\lambda=\lambda_{B})&=&2\pi\lambda_{B}^{3}\left(\frac{\pi}{3}+\frac{\sqrt{3}}{2}\right),
\label{a3n123}
\eea
consistent with the total three dimensional volume $V(S^{3};\lambda=\lambda_{B})=2\pi^{2}\lambda_{B}^{3}$ 
in (\ref{volumeblambda}), which is applicable to the neutron and the proton. 
Following the scanning algorithm discussed in the previous section, we notice that 
$V_{n}^{I}(S^{3};\lambda=\lambda_{B})+V_{n}^{III}(S^{3};\lambda=\lambda_{B})$ is located in the inner portion of the neutron, while 
$V_{n}^{II}(S^{3};\lambda=\lambda_{B})$ is situated in the outer portion of the neutron. Similar arguments can be applied to the 
fractional neutron charges in (\ref{qn123}) as shown in Fig.~\ref{chargep}(b) and Fig.~\ref{hyper}(b). 

Now it seems appropriate to comment on the numerical analysis of the neutron in the HSM. The predicted three 
dimensional volumes for the fractional neutron distributions in (\ref{a3n123}) and the corresponding total volume in (\ref{volumeblambda}) 
are listed in Table~II. Note that in the neutron hypersphere soliton the mass and charge fill the 
enhanced $S^{3}$ volume comparing to the corresponding ordinary $R^{3}$ one, to yield
\beq
\frac{V_{n}(S^{3};\lambda=\lambda_{B})}{V_{n}(R^{3};\lambda=\lambda_{B})}=\left\{\begin{array}{ll}
4\left(\frac{\pi}{3}-\frac{\sqrt{3}}{4}\right)=2.457,~~~~{\rm for~volume~inside~the~core,}\\
6\left(\frac{\pi}{3}+\frac{\sqrt{3}}{2}\right)=11.479,~~~{\rm for~volume~outside~the~core,}
\end{array}\right.
\label{ratiovolumes}
\eeq
implying that, in the neutron in the HSM, the volume ratio outside the core is much more enhanced than 
that inside the core. Note also that the volume inside the neutron core in the $S^{3}$ hypersphere originates from 
{\it two regions} $V_{n}^{I}(S^{3};\lambda=\lambda_{B})$ and $V_{n}^{III}(S^{3};\lambda=\lambda_{B})$ shown 
in Fig.~\ref{hyper}(b), while that inside the core in $R^{3}$ comes from {\it one region} 
$V_{n}^{inside}(R^{3};\lambda=\lambda_{B})$ associated with the ordinary solid sphere of radius 
$\frac{\sqrt{3}}{2}\lambda_{B}$. Moreover, the ratio of total volumes listed in 
Table~II reproduces the characteristic ratio $\chi=\frac{3\pi}{2}=4.712$ in (\ref{chiratio}).

\begin{table}[t]
\caption{The numerical predictions of the three dimensional volumes for the fractional neutron distributions. 
Here the volume inside (outside) the neutron core is positively (negatively) charged as shown in Fig.~\ref{hyper}(b).}
\begin{center}
\begin{tabular}{lll}
\hline
Volume type  &~~~~$S^{3}$ &~~~$R^{3}$\\
\hline
Volume inside the core  &~~~~$5.188~{\rm fm}^{3}$ &~~~$2.112~{\rm fm}^{3}$\\
Volume outside the core &~~~~$8.081~{\rm fm}^{3}$ &~~~$0.704~{\rm fm}^{3}$\\
Total volume &~~~$13.269~{\rm fm}^{3}$ &~~~$2.816~{\rm fm}^{3}$\\
\hline
\end{tabular}
\end{center}
\label{tablevolume}
\end{table}

\section{Hopf-links and topological M\"obius strips in HSM}
\setcounter{equation}{0}
\renewcommand{\theequation}{\arabic{section}.\arabic{equation}}

In this section, we investigate Hopf-links of the M\"obius strips in the HSM. To accomplish this, we first proceed to study 
baryon phenomenology by using the Skyrmion Lagrangian density in (\ref{lagtot}) and the three metric on $S^{3}$ in (\ref{metrics3}). 
For the minimum energy of the soliton we can take the hedge ansatz $U_{0}=e^{i\sigma_{a}\hat{x}_{a}f(\mu)}$ 
where $\sigma_{a}$ $(a=1,2,3)$ is the Pauli matrix. In the hypersphere soliton, with the hedgehog ansatz and the collective coordinates $A(t)\in {\rm SU(2)}$, 
the chiral field can be given by 
\beq
U=A(t)U_{0}A^{\dagger}(t)=e^{i\sigma_{a}A_{ab}\hat{x}_{b}f(\mu)},
\label{uxt}
\eeq
where 
\beq
A_{ab}=\frac{1}{2}{\rm tr}(\sigma_{a}A\sigma_{b}A^{\dagger}).
\label{aab}
\eeq
In the first class Dirac quantization~\cite{hong21} possessing the WOC~\cite{hong21,lee}, we construct the energy spectrum of the form
\beq
H=E+\frac{1}{2{\cal I}}\left[I(I+1)+\frac{1}{4}\right],
\label{spectromwoc}
\eeq
where $I$ is the isospin quantum number having $1/2$ ($3/2$) for the nucleon (delta baryon) and 
${\cal I}$ is the moment of inertia of the hypersphere soliton given by
\beq
{\cal I}=\frac{2f_{\pi}^{2}}{3}\int_{S^{3}}dV_{B}\left(1+\left(\frac{df}{d\mu}\right)^{2}+\frac{\sin^{2}f}{\sin^{2}\mu}\right)\sin^{2}f.
\label{cali}
\eeq

Next, in the hypersphere soliton on $S^{3}$, we obtain the soliton energy~\cite{manton1,hong98plb,hong21}
\beq
E=\frac{f_{\pi}}{e}\left[2\pi L\int_{0}^{\pi}d\mu\sin^{2}\mu\left(\left(\frac{d f}{d\mu}
    +\frac{1}{L}\frac{\sin^{2}f}{\sin^{2}\mu}\right)^{2}+2\left(\frac{1}{L}\frac{d f} 
     {d\mu}+1\right)^{2}\frac{\sin^{2}f}{\sin^{2}\mu}\right)
+6\pi^{2}B\right],
\label{e2}
\eeq
where $L=ef_{\pi}\lambda$ ($0\le L<\infty$) is a radius expressed in dimensionless units, and 
$f(\mu)$ is the profile function for the hypersphere soliton and it satisfies $f(0)=\pi$ and $f(\pi)=0$ for 
unit baryon number. Here $B$ is a conserved topological charge which is a baryon number in the HSM 
given by~\cite{hong98plb}
\beq
B=\int_{S^{3}}dV_{B}B^{0}=-\frac{2}{\pi}\int_{0}^{\pi} d\mu \sin^{2} f \frac{df}{d\mu},
\label{bfmu}
\eeq
where $B^{0}$ is a baryon number current given by (\ref{b0bidef1}). Exploiting (\ref{spectromwoc})--(\ref{bfmu}) and the 
identity map condition $f(\mu)=\pi-\mu$, 
the nucleon mass $M_{N}$ ($I=1/2$) and delta baryon mass $M_{\Delta}$ ($I=3/2$) are given by~\cite{hong21}
\begin{equation}
M_{N}=ef_{\pi}\left(\frac{6\pi^{2}}{e^{2}}+\frac{e^{2}}{6\pi^{2}}\right),~~~
M_{\Delta}=ef_{\pi}\left(\frac{6\pi^{2}}{e^{2}}+\frac{2e^{2}}{3\pi^{2}}\right).
\label{mnmdelta}
\end{equation}

Now we have brief comments on the geometrical Hopf-links of the M\"obius strips in the HSM.
First, we consider a circle $S^{1}(\subset S^{3})$ on the hypersphere. 
For the space manifolds $S^{3}$ and $S^{2}$, we have the homotopy 
group~\cite{toda,manton2,schwarz}
$\Pi_{3}(S^{2})={\mathbf Z}$. Note that in the homotopy group, 
the target manifold $S^{2}$ is a geometrical space. Due to this homotopy group, 
we have an associated integer topological charge, namely the 
Hopf charge~\cite{manton2}. Here the Hopf charge is intrinsically distinct from the topological charge $B$ associated with 
$\Pi_{3}(S^{3})={\mathbf Z}$. In other words, we cannot construct any Hopf charge having a form similar to $B$ in (\ref{bfmu}). 
Instead, we have the preimage of a point on the target manifold $S^{2}$ which is a closed loop in $S^{3}$~\cite{manton2,bott}.

Second, it is shown in the Hopf fibration~\cite{manton2,bott,hopf,ryder} that the closed loop is a twist circle $S^{1}(\subset S^{3})$. 
For instance, using the homology groups $H_{1}(S^{3})=0$ and $H_{1}(S^{2}\times S^{1})={\mathbf Z}$, $S^{3}$ on which the closed loop $S^{1}(\subset S^{3})$ 
is embedded, is shown to be locally (not globally) equivalent to $S^{2}\times S^{1}$ where $S^{1}$ is an ordinary circle~\cite{ryder}. 
This aspect implies that there exists the twist circle on the hypersphere in the HSM.
Moreover, the preimages in $S^{3}$ decompose the hypersphere into a continuous family of circles, and two distinct circles are linked in $S^{3}$ 
and form~\cite{manton2,bott} the Hopf link~\cite{kauffman,baez}.

Next, in order to investigate further the topological structure of the twist circle in the HSM,
we first consider a toy model of 
the (1+1) dimensional sine-Gordon soliton whose Lagrangian is given in 
appropriate length and energy units as follows~\cite{skyrme61b,raja,proceeding,manton2}
\beq
L_{SG}=\int dx \left[\frac{1}{2}\pa_{\mu}\phi\pa^{\mu}\phi+\cos\phi-1\right],
\label{sglag}
\eeq
from which we obtain the field equation of the form
\beq
\Box\phi+\sin \phi=0.
\eeq
Note that the sine-Gordon field equation satisfies discrete symmetry 
$\phi (x,t)\rightarrow \phi (x,t)+2N\pi$ ($N=...-2, -1, 0, 1, 2, ...$). 
The topological charge for a static solution of the sine-Gordon kink is then given by 
$Q_{top}=\frac{1}{2\pi}[\phi(\infty)-\phi(-\infty)]$. 
We now have the static soliton solution of a monotonically increasing {\it profile}: 
$\phi (x)=4\tan^{-1}\exp x$ $(-\infty<x<\infty)$ having $\phi(-\infty)=0$, $\phi(0)=\pi$ and $\phi(\infty)=2\pi$ 
for the case of $Q_{top}=1$ of the kink, to imply that the sine-Gordon kink solution represents 
infinite M\"obius strip feature~\cite{raja,proceeding}. 
Moreover the sine-Gordon kink state is identifiable as a fermion~\cite{raja}. Exploiting $B$ in (\ref{bfmu}), we 
have the {\it profile function solution} $f(\mu)=\pi-\mu$ $(0\le\mu\le\pi)$ possessing $f(0)=\pi$ and $f(\pi)=0$ 
for the hypersphere soliton with a topological charge 
$B=1$~\cite{hong98plb,manton2}. This implies that in $S^{1}(\subset S^{3})$ defined in the range $0\le\mu\le\pi$ 
we have the fermionic M\"obius strip feature similar to that in the sine-Gordon kink. Note that the topological feature in $S^{3}$ is mainly
described in terms of $S^{1}$, since we have no nontrivial topological property in $S^{2}$ in the hypersphere $S^{3}(=S^{2}\times S^{1})$. 
Note also that this M\"obius strip topological structure corresponds to the twist circle one in the HSM.

Now, in order to study the physical meaning of the angle $\mu$ which is related with the range of the 
twist circle $S^{1}(\subset S^{3})$, we investigate the phenomenological aspects in 
the neutron charge density $\rho_{n}(\mu)$ depicted in Fig.~\ref{chargep}(b). Here the positive (negative) charge 
density is located in the range $0\le\mu\le\frac{\pi}{3}$ and $\frac{2\pi}{3}\le\mu\le\pi$ 
(in $\frac{\pi}{3}\le\mu\le\frac{2\pi}{3}$), to produce a cyclic sequence
\beq
I\left(0\le\mu\le\frac{\pi}{3};~Q_{I}\right)\rightarrow II\left(\frac{\pi}{3}\le\mu\le\frac{2\pi}{3};~Q_{II}\right)\rightarrow 
III\left(\frac{2\pi}{3}\le\mu\le\pi;~Q_{III}\right),
\label{sequence}
\eeq
corresponding to a twist circle associated with the M\"obius strip structure of the neutron.
Note that we have the phase 
transition of the charges at the angles $\mu=\frac{\pi}{3}$ and $\mu=\frac{2\pi}{3}$ in the neutron, which has 
the twist circle in $S^{3}$. In contrast, the proton has no phase transition of the charges 
even though it has the topological feature similar to that of the neutron.

Next we have comments on the classification of the boson and fermion in terms of the strip structures. The fermion performs $4\pi$ 
rotation to return its initial situation according to the relativistic quantum 
mechanics~\cite{bjorken64,hong22massive}. This 
fermion characteristic is elucidated by using $4\pi$ rotation on the M\"obius strip 
described above. The baryon inner structure is thus successfully explained on a hypersphere $S^{3}$ possessing 
the M\"obius strip type twist circle. In contrast, the boson performs $2\pi$ rotation to return its starting situation. The boson feature is 
then explained by exploiting $2\pi$ rotation on the ordinary strip, implying that we do not need the 
hypersphere geometry associated with the M\"obius strip type twist circle in revealing the boson inner structure.

\section{Conclusions}
\setcounter{equation}{0}
\renewcommand{\theequation}{\arabic{section}.\arabic{equation}}
\label{conclusion}

In summary, we have investigated the geometrical and topological bindings of the neutron and proton, by exploiting 
the foliation in the HSM. To accomplish this, we have studied three dimensional hypersphere $S^{3}$ of 
radius $\lambda$ which is
delineated by $(\mu,\theta,\phi)$, by exploiting a 
projection from the northern hemi-hypersphere $S^{3}_{+}$ to the three dimensional equatorial solid sphere 
$E^{3}$ which possesses the center of $S^{3}$. Introducing the so-called scanning algorithm, we have found 
similarities between the spherical shell foliation leave on the curved manifold $S^{3}_{+}$ and that on $E^{3}$. 
Here both spherical shell foliation leaves on $S^{3}_{+}$ and $E^{3}$ have been shown to share their 
corresponding centers, differently from a toy model of the spherical shell $S^{2}$ where the $S^{1}$ foliation leaves do not share 
the center of $S^{2}$. Making use of the foliation leaves on $S^{2}$ of radius $\lambda\sin\mu$ to scan the curved manifold $S^{3}$, 
we have constructed the volume $V(S^{3})=2\pi^{2}\lambda^{3}$, which is consistent with the result from the pure geometrical approach without resorting to the 
scanning algorithm. Next, exploiting the geometry on the three dimensional volume of the hypersphere soliton, we have formulated the 
three dimensional fractional volumes to explicitly yield the corresponding numerical predictions. 
We then have found that the characteristic ratio of the hypersphere volume $V(S^{3};\lambda=\lambda_{B})$ to 
the corresponding solid sphere one $V(R^{3};\lambda=\lambda_{B})$ is given by the geometrical invariant implying that 
the baryon charges are hyper-compactly confined inside this hypersphere volume. We also have clarified the physical meaning of 
$\mu$ in $S^{3}$ of radius $\lambda$. Constructing the charge density profiles of nucleons in terms of the angle $\mu$ 
in $S^{3}$, we have found that the proton and neutron charge densities are spherically symmetric with 
the nontrivial distribution structures without leaking out from the soliton. In particular, in the neutron 
charge distribution, the positive (negative) charge has been shown to be confined inside (outside) its core. 
Explicitly we have obtained the charge fractions in the neutron. We next have found the phase transition of 
the charges at the angles $\mu=\frac{\pi}{3}$ and $\mu=\frac{2\pi}{3}$ 
in the neutron, which possesses the topology of two Hopf-linked~\cite{manton2,bott,kauffman,baez} 
M\"obius strip type twist circles in $S^{3}$. However, the proton has been shown to have no phase transition of the charges, 
even though it has the topological structure similar to that of the neutron. 
This topological property of the HSM defined on the hyper-compact manifold $S^{3}$ 
is one of the main points of our paper. Next we have briefly classified the boson and fermion via 
the strip structures. Note that we have the algebraic binding of baryons associated with the 
BPS lower bound of hypersphere soliton.
Note also that the theoretical predictions for the physical quantities in the HSM are in good agreement with 
the corresponding experimental data as shown in Table~I. For instance, the prediction for $\mu_{\Delta^{++}}=5.69$ 
in the HSM~\cite{hong21} is within the corresponding experimental value 
$\mu_{\Delta^{++}}^{\rm exp}=4.7-6.7$~\cite{boss}, even though the HSM has nontrivial geometrical and topological structures as shown in this paper.

\acknowledgments{The author would like to thank the anonymous referees for helpful comments. He 
was supported by Basic Science Research Program through the National Research Foundation of Korea 
funded by the Ministry of Education, NRF-2019R1I1A1A01058449.}\\ \\

\noindent
{\bf Data availability statement}: All data that support the findings of this study are included within the article (and any supplementary files).\\ \\

\appendix
\section{Nucleon charges, charge radii and magnetic moments}
\label{zetachapter}
\setcounter{equation}{0}
\renewcommand{\theequation}{A.\arabic{equation}}

\subsection{Formalism for nucleon charges and electric mean square charge radii in HSM}

Now we study the formalism for the nucleon charges in the HSM by starting with the charge operator
\beq
Q=\int_{S^{3}} dV_{B} J^{0},
\label{defq}
\eeq 
where $J^{0}=\frac{1}{2}B^{0}+J_{V}^{03}$ is the time component of the electromagnetic current $J^{\mu}=\frac{1}{2}B^{\mu}+J_{V}^{\mu 3}$. 
Before formulating the physical quantities, we digress to comment on the electromagnetic current $J^{\mu}$ in the HSM. First, $B^{\mu}$ is the 
topological current given by~\cite{skyrme61}
\beq
B^{\mu}=\frac{1}{24\pi^{2}}\epsilon^{\mu\nu\rho\sigma}{\rm tr}(L_{\nu}L_{\rho}L_{\sigma}),
\label{bmudef}
\eeq
where $L_{\mu}=U^{\dagger}\pa_{\mu}U$ and $\epsilon^{1230}=1$. Inserting the chiral field $U$ in (\ref{uxt}) into $B^{\mu}$ in (\ref{bmudef}), 
in the HSM we construct the time and spatial components of the topological current $B^{\mu}$ as follows 
\bea
B^{0}&=&-\frac{e^{3}f_{\pi}^{3}}{2\pi^{2}}\frac{\sin^{2}f}{\sin^{2}\mu}\frac{df}{d\mu},\label{b0bidef1}\\
B^{i}&=&\frac{e^{2}f_{\pi}^{2}}{\pi^{2}}\frac{\sin^{2}f}{\sin\mu}\frac{df}{d\mu}\epsilon_{iab}\hat{x}_{a}A_{b},
\label{b0bidef2}
\eea
where $A_{a}$ is defined as
\beq
A_{a}=-\frac{i}{2} {\rm tr} (\sigma_{a}A^{\dagger}\partial_{0}A).
\label{defaa}
\eeq
Note that, in formulating $B^{0}$ in (\ref{b0bidef1}), we can use the hedgehog solution $U_{0}$ effectively, since we have only the spatial 
derivatives in the time component of $B^{\mu}$ in (\ref{bmudef}).

Second, $J_{V}^{\mu 3}$ is the vector current obtainable by applying the Noether method to 
the Skyrmion Lagrangian density in (\ref{lagtot}). To be specific, in order to construct $J_{V}^{\mu 3}$, we consider infinitesimal 
isospin transformation,
\beq
U\rightarrow SUS^{\dagger},
\label{infinitesimal}
\eeq
where $S=I-\frac{i}{2}\varepsilon_{a}\sigma_{a}$ with $\varepsilon_{a}$ being the infinitesimal transformation parameter. 
Making use of the definition $\delta U=SUS^{\dagger}-U$ and the Skyrmion Lagrangian density in (\ref{lagtot}), we construct
\beq
\delta {\cal L}=(\pa_{\mu}\varepsilon_{a})J_{V}^{\mu a},
\label{deltalagsk}
\eeq
where in the HSM the vector currents are given by
\bea
J_{V}^{03}&=&-2f_{\pi}^{2}\left(1+\left(\frac{df}{\d\mu}\right)^{2}+\frac{\sin^{2}f}{\sin^{2}\mu}\right)\sin^{2}f A_{a}(\delta_{ab}
-\hat{x}_{a}\hat{x}_{b})A_{3b},\label{defbj0}\\
J_{V}^{i3}&=&ef_{\pi}^{3}\frac{\sin^{2}f}{\sin\mu}\left(1+\left(\frac{df}{\d\mu}\right)^{2}+\frac{\sin^{2}f}{\sin^{2}\mu}\right)\epsilon_{iab}\hat{x}_{a}A_{3b},
\label{defbj}
\eea
where $A_{a}$ and $A_{ab}$ are given by (\ref{defaa}) and (\ref{aab}), respectively. Note that we have 
the solid angle integration identity of $A_{a}$ and $A_{ab}$,
\beq
\int d\Omega A_{a}(\delta_{ab}-\hat{x}_{a}\hat{x}_{b})A_{3b}=-\frac{4\pi I_{3}}{3{\cal I}},
\label{idaa}
\eeq
where $I_{3}$ is the isospin operator and ${\cal I}$ is the moment of the inertia in (\ref{cali}).

Next, keeping in mind the baryon number current $B^{0}$ in (\ref{b0bidef1}) and the third component of vector one $J_{V}^{0 3}$ 
in (\ref{defbj0}), we proceed to construct the charge operator $Q$ in (\ref{defq}) in terms of the charge densities and the isospin quantum number
\beq
Q=\int_{0}^{\pi}d\mu\left(\frac{1}{2}\rho_{I=0}(\mu)+I_{3}\rho_{I=1}(\mu)\right),
\label{qrho}
\eeq 
where the isoscalar and isovector charge densities are given by~\cite{hong98plb}
\bea
\rho_{I=0}(\mu)&=&-\frac{2}{\pi}\sin^{2}f\frac{df}{d\mu},\label{rho0}\\
\rho_{I=1}(\mu)&=&\frac{8\pi}{3\tilde{\cal I}}\left(1+\left(\frac{df}{d\mu}\right)^{2}+\frac{\sin^{2}f}{\sin^{2}\mu}\right)\sin^{2}\mu\sin^{2}f,
\label{rho12s}
\eea
with $\tilde{\cal I}$ being a dimensionless variable defined as $\tilde{\cal I}=e^{3}f_{\pi}{\cal I}$. Note that the charge operator in (\ref{qrho}) yields 
\beq
Q_{p,n}=\frac{e^{3}f_{\pi}^{3}}{4\pi^{2}}\int_{S^{3}} dV_{B}\left(1\pm \frac{4}{3}\sin^{2}\mu\right),
\label{qrhopn}
\eeq 
to reproduce $Q_{p,n}$ in (\ref{pncharge2})
\beq
Q_{p,n}=\int_{0}^{\pi}d\mu\rho_{p,n}(\mu),
\label{pncharge22}
\eeq 
and the charge density profiles $\rho_{p}(\mu)$ and $\rho_{n}(\mu)$ in (\ref{rhopn})
\beq
\rho_{p,n}(\mu)=\frac{1}{2}\left(\rho_{I=0}(\mu)\pm \rho_{I=1}(\mu)\right)=\frac{1}{\pi}\sin^2\mu \left(1\pm \frac{4}{3}\sin^2 \mu\right).
\label{rhopn2}
\eeq
Note also that, after integrating $Q_{p,n}$ in (\ref{pncharge22}) over the angle $\mu$, we find $Q_{p}=1$ and $Q_{n}=0$. This aspect implies that 
the baryon number current $B^{0}$ in (\ref{b0bidef1}) and the third component of vector one $J_{V}^{03}$ in (\ref{defbj0}) and the ensuing 
isoscalar and isovector charge densities $\rho_{I=0}(\mu)$ and $\rho_{I=0}(\mu)$ in (\ref{rho0}) and (\ref{rho12s}) are well constructed.

Next we study the formalism for the electric isoscalar and isovector mean square charge radii in the HSM. 
Exploiting the isoscalar charge density in (\ref{rho0}), in the HSM we construct the electric isoscalar mean square charge radius given by
\beq
\langle r^{2}\rangle_{E,I=0}=\int_{0}^{\pi}d\mu(\lambda_{B}\sin\mu)^{2}\rho_{I=0}(\mu).
\label{r2i00}
\eeq
Combining (\ref{rho0}) and (\ref{r2i00}), we end up with
\beq
\langle r^{2}\rangle_{E,I=0}=\frac{3}{4}\frac{1}{e^{2}f_{\pi}^{2}}.
\label{r2i0f}
\eeq
In the HSM the electric isovector mean square charge radius is given by
\beq
\langle r^{2}\rangle_{E,I=1}=\int_{0}^{\pi}d\mu(\lambda_{B}\sin\mu)^{2}\rho_{I=1}(\mu).
\label{r2i0}
\eeq
Exploiting (\ref{rho12s}) and (\ref{r2i0}), we obtain
\beq
\langle r^{2}\rangle_{E,I=1}=\frac{5}{6}\frac{1}{e^{2}f_{\pi}^{2}}.
\label{r2i1f}
\eeq
Making use of (\ref{r2i0f}) and (\ref{r2i1f}), we reproduce the electric isoscalar and isovector mean charge radii in (\ref{radii4}) as follows
\beq
\langle r^{2}\rangle_{E,I=0}^{1/2}=\frac{\sqrt{3}}{2}\frac{1}{ef_{\pi}},~~~\langle r^{2}\rangle_{E,I=1}^{1/2}=\sqrt{\frac{5}{6}}\frac{1}{ef_{\pi}}.
\eeq
The proton and neutron mean square charge radii in the HSM are given by 
\beq
\langle r^{2}\rangle_{p,n}=\int_{0}^{\pi}d\mu(\lambda_{B}\sin\mu)^{2}\rho_{p,n}(\mu).
\label{r2pn}
\eeq
Inserting $\rho_{p,n}(\mu)$ in (\ref{rhopn2}) into (\ref{r2pn}), we construct
\beq
\langle r^{2}\rangle_{p}=\frac{19}{24}\frac{1}{e^{2}f_{\pi}^{2}},~~~\langle r^{2}\rangle_{n}=-\frac{1}{24}\frac{1}{e^{2}f_{\pi}^{2}}.
\label{r2pn2}
\eeq

\subsection{Formalism for magnetic moments and magnetic mean square charge radii in HSM}

Now we investigate the formalism for the magnetic moments and the ensuing magnetic mean square charge radii in the HSM. 
To do this, we start with the magnetic moment operator
\beq
\mu^{i}=\frac{1}{2}\int_{S^{3}}dV_{B}\epsilon_{ijk}(\lambda_{B}\sin\mu~\hat{x}_{j})J^{k}
\label{magmom}
\eeq 
where $J^{i}=\frac{1}{2}B^{i}+J_{V}^{i3}$ is the spatial component of the electromagnetic current $J^{\mu}$.

Inserting $B^{i}$ in (\ref{b0bidef2}) and $J_{V}^{i3}$ in (\ref{defbj}) into $\mu^{i}$ in (\ref{magmom}), we construct the magnetic moment operator
\beq
\mu^{i}=\frac{1}{6{\cal I}}\langle r^{2}\rangle_{E,I=0}S_{i}-\frac{1}{2}{\cal I}A_{3i},
\label{magmom2}
\eeq
where the soliton spin operator is given by $S_{i}=2{\cal I}A_{i}$. Making use of 
(\ref{magmom2}), the proton magnetic moment is given by~\cite{hong98plb}
\beq
\mu_{p}=2M_{N}\langle p~1/2|\mu^{3}|p~1/2\rangle=2M_{N}\left(\frac{1}{12{\cal I}}\langle r^{2}\rangle_{E,I=0}+\frac{\cal I}{6}\right)
=\frac{2M_{N}}{ef_{\pi}}\left(\frac{1}{48}\frac{e^{2}}{\pi^{2}}+\frac{1}{2}\frac{\pi^{2}}{e^{2}}\right),
\label{protonmag}
\eeq
where $M_{N}$ is given by (\ref{mnmdelta}) and we have used the identities $\langle p~1/2|S_{3}|p~1/2\rangle=1/2$ and $\langle p~1/2|A_{33}|p~1/2\rangle=-1/3$. 
Similarly the neutron magnetic moment is given by~\cite{hong98plb}
\beq
\mu_{n}=2M_{N}\langle n~1/2|\mu^{3}|n~1/2\rangle=2M_{N}\left(\frac{1}{12{\cal I}}\langle r^{2}\rangle_{E,I=0}-\frac{\cal I}{6}\right)
=\frac{2M_{N}}{ef_{\pi}}\left(\frac{1}{48}\frac{e^{2}}{\pi^{2}}-\frac{1}{2}\frac{\pi^{2}}{e^{2}}\right).
\label{neutronmag}
\eeq
Next we find the delta baryon magnetic moment given by~\cite{hong98plb}
\beq
\mu_{\Delta^{++}}=2M_{N}\langle\Delta^{++}~3/2|\mu^{3}|\Delta^{++}~3/2\rangle=2M_{N}\left(\frac{1}{4{\cal I}}\langle r^{2}\rangle_{E,I=0}+\frac{3{\cal I}}{10}\right)
=\frac{2M_{N}}{ef_{\pi}}\left(\frac{1}{16}\frac{e^{2}}{\pi^{2}}+\frac{9}{10}\frac{\pi^{2}}{e^{2}}\right),
\label{deltamag}
\eeq
where we have used the identities $\langle \Delta^{++}~3/2|S_{3}|\Delta^{++}~3/2\rangle=3/2$ and 
$\langle \Delta^{++}~3/2|A_{33}|\Delta^{++}~3/2\rangle=-3/5$

Now we study the normalized magnetic moment densities by splitting (\ref{magmom2}) into the isoscalar part and isovector one,
\bea
\mu^{i}_{M,I=0}&=&\int_{0}^{\pi}d\mu\left(-\frac{1}{3\pi{\cal I}}(\lambda_{B}\sin\mu)^{2}\sin^{2}f\frac{df}{d\mu}S_{i}\right),\nn\\
\mu^{i}_{M,I=1}&=&\int_{0}^{\pi}d\mu\left(-\frac{4\pi f_{\pi}}{3e}(\lambda_{B}\sin\mu)^{2}\sin^{2}f\left(1+\left(\frac{df}{d\mu}\right)^{2}
+\frac{\sin^{2}f}{\sin^{2}\mu}\right)A_{3i}\right),
\label{twoparts}
\eea
from which we define the normalized isoscalar and isovector magnetic moment densities
\bea
\rho_{M,I=0}&=&\frac{-\frac{1}{3\pi{\cal I}}(\lambda_{B}\sin\mu)^{2}\sin^{2}f\frac{df}{d\mu}S_{i}}
{\mu^{i}_{M,I=0}}
=\frac{(\lambda_{B}\sin\mu)^{2}\rho_{I=0}(\mu)}{\langle r^{2}\rangle_{E,I=0}},\label{twodensities1}\\
\rho_{M,I=1}&=&\frac{-\frac{4\pi f_{\pi}}{3e}(\lambda_{B}\sin\mu)^{2}\sin^{2}f\left(1+\left(\frac{df}{d\mu}\right)^{2}
+\frac{\sin^{2}f}{\sin^{2}\mu}\right)A_{3i}}{\mu^{i}_{M,I=1}}
=\rho_{I=1}.
\label{twodensities}
\eea
Note that we have used the relation $\int_{0}^{\pi}d\mu \rho_{I=1}(\mu)=1$ in (\ref{twodensities}). 
Making use of (\ref{twodensities1}) and (\ref{twodensities}), we construct the magnetic isoscalar and isovector 
mean square charge radii as follows~\cite{hong98plb}
\bea
\langle r^{2}\rangle_{M,I=0}&=&\int_{0}^{\pi}d\mu(\lambda_{B}\sin\mu)^{2}\rho_{M,I=0}(\mu)
=\frac{\langle r^{4}\rangle_{E,I=0}}{\langle r^{2}\rangle_{E,I=0}}=\frac{5}{6}\frac{1}{e^{2}f_{\pi}^{2}},\nn\\
\langle r^{2}\rangle_{M,I=1}&=&\int_{0}^{\pi}d\mu(\lambda_{B}\sin\mu)^{2}\rho_{M,I=1}(\mu)
=\langle r^{2}\rangle_{E,I=1}=\frac{5}{6}\frac{1}{e^{2}f_{\pi}^{2}},
\label{r2m0m1def}
\eea
where we have exploited the following integral
\beq
\langle r^{4}\rangle_{E,I=0}=\int_{0}^{\pi}d\mu(\lambda_{B}\sin\mu)^{4}\rho_{I=0}(\mu)=\frac{5}{8}\frac{1}{e^{4}f_{\pi}^{4}}.
\label{r4m0def}
\eeq

Next we define the magnetic proton and neutron mean square charge radii $\langle r^{2}\rangle_{M,p}$ and $\langle r^{2}\rangle_{M,n}$ 
in terms of the corresponding magnetic isoscalar and isovector 
mean square charge radii in (\ref{r2m0m1def}), and the proton and neutron magnetic moments in (\ref{protonmag}) and (\ref{neutronmag})
\bea
\mu_{p}\langle r^{2}\rangle_{M,p}&=&2M_{N}\left(\frac{1}{12{\cal I}}\langle r^{2}\rangle_{E,I=0}\langle r^{2}\rangle_{M,I=0}
+\frac{\cal I}{6}\langle r^{2}\rangle_{M,I=1}\right),\nn\\
\mu_{n}\langle r^{2}\rangle_{M,n}&=&2M_{N}\left(\frac{1}{12{\cal I}}\langle r^{2}\rangle_{E,I=0}\langle r^{2}\rangle_{M,I=0}
-\frac{\cal I}{6}\langle r^{2}\rangle_{M,I=1}\right),
\label{r2mpn}
\eea
from which we find
\bea
\mu_{p}\langle r^{2}\rangle_{M,p}+\mu_{n}\langle r^{2}\rangle_{M,n}&=&(\mu_{p}+\mu_{n})\langle r^{2}\rangle_{M,I=0},\nn\\
\mu_{p}\langle r^{2}\rangle_{M,p}-\mu_{n}\langle r^{2}\rangle_{M,n}&=&(\mu_{p}-\mu_{n})\langle r^{2}\rangle_{M,I=1}.
\label{idmupmun}
\eea
Here we have used the identities $\mu_{p}+\mu_{n}=\frac{1}{3{\cal I}}M_{N}\langle r^{2}\rangle_{E,I=0}$ and 
$\mu_{p}-\mu_{n}=\frac{2}{3}M_{N}{\cal I}$ which are obtainable from (\ref{protonmag}) and (\ref{neutronmag}). Shuffling the equations in (\ref{idmupmun}), we arrive at
\bea
\langle r^{2}\rangle_{M,p}&=&\frac{1}{2\mu_{p}}\left((\mu_{p}+\mu_{n})\langle r^{2}\rangle_{M,I=0}+(\mu_{p}-\mu_{n})\langle r^{2}\rangle_{M,I=1}\right),\nn\\
\langle r^{2}\rangle_{M,n}&=&\frac{1}{2\mu_{n}}\left((\mu_{p}+\mu_{n})\langle r^{2}\rangle_{M,I=0}-(\mu_{p}-\mu_{n})\langle r^{2}\rangle_{M,I=1}\right).
\label{r2mpn2}
\eea
Inserting (\ref{r2m0m1def}) into (\ref{r2mpn2}), we 
end up with
\beq
\langle r^{2}\rangle_{M,p}=\langle r^{2}\rangle_{M,n}=\frac{5}{6}\frac{1}{e^{2}f_{\pi}^{2}},
\label{r2mpn22}
\eeq
which reproduce $\langle r^{2}\rangle_{M,p}^{1/2}$ and $\langle r^{2}\rangle_{M,n}^{1/2}$ in (\ref{radii4}).



\begin{thebibliography}{99}
\bibliographystyle{unsrt}
\bibitem{di} P.A.M. Dirac, {\it Lectures on Quantum Mechanics} (Yeshiva University Press, 1964).
\bibitem{hong15} S.T. Hong, {\it BRST Symmetry and de Rham Cohomology} (Springer, 2024) (https://doi.org/10.1007/978-981-97-0960-1).
\bibitem{anw83} G.S. Adkins, C.R. Nappi, E. Witten, {\it Static properties of nucleons in the Skyrme model}, Nucl. Phys. B {\bf 228}, 552 (1983).
\bibitem{skyrme61} T.H.R. Skyrme, {\it A nonlinear field theory}, Proc. R. Soc. Lond. A {\bf 260}, 127 (1961).
\bibitem{hong98plb} S.T. Hong, {\it The static properties of Skyrmions on a hypersphere}, Phys. Lett. B {\bf 417}, 211 (1998).
\bibitem{manton1} N.S. Manton, P.J. Ruback, {\it Skyrmions in flat space and curved space}, Phys. Lett. B {\bf 181}, 137 (1986).
\bibitem{hong21} S.T. Hong, {\it Dirac quantization and baryon intrinsic frequencies in hypersphere soliton model}, 
Nucl. Phys. B {\bf 973}, 115611 (2021), arXiv:2105.11456.
\bibitem{foli} A. Candel, L. Conlon, {\it Foliations I} (American Mathematical Society, 2000).
\bibitem{stern33} R. Frisch, O. Stern, {\it \"{U}ber die magnetische ablenkung von wasserstoffmolekulen und das magnetische 
moment des protons. I}, Z. Physik {\bf 85}, 4 (1933).
\bibitem{cg} S. Coleman, S.L. Glashow, {\it Electrodynamic properties of baryons in the unitary symmetry scheme}, Phys. Rev. Lett. {\bf 6}, 423 (1961).
\bibitem{boss} A. Bosshard et al., {\it Analyzing power in pion proton bremsstrahlung, and the Delta++ (1232) magnetic moment}, Phys. Rev. D {\bf 44}, 1962 (1991).
\bibitem{manton2} N.S. Manton and P. Sutcliffe, {\it Topological Solitons} (Cambridge University Press, 2004) (https://doi.org/10.1017/CBO9780511617034).
\bibitem{bott} R. Bott, L.W. Tu, {\it Differential Forms in Algebraic Topology} (Springer, 1982).
\bibitem{kauffman} L.H. Kauffman, {\it On Knots} (Princeton University Press, 1987). 
\bibitem{baez} J. Baez, J.P. Muniain, {\it Gauge Fields, Knots and Gravity} (World Scientific, 1994).
\bibitem{oka} A. Oka, A. Hosaka, {\it Nucleons as Skyrmions}, Ann. Rev. Nucl. Part. Sci. {\bf 42}, 333 (1992).
\bibitem{liu87} G.S. Adkins, {\it Chiral Solitons} (World Scientific, 1987), Editor: K.F. Liu.
\bibitem{lee} T.D. Lee, {\it Particle Physics and Introduction to Field Theory} (Harwood Academic Publishers, 1981) p. 477.
\bibitem{toda} H. Toda, {\it Composition Methods in Homotopy Groups of Spheres} (Princeton University Press, 1962).
\bibitem{schwarz} A.S. Schwarz, {\it Quantum Field Theory and Topology} (Springer, 1993).
\bibitem{hopf} H. Hopf, {\it \"Uber die abbildungen der dreidimensionalen sph\"are auf die kugelfl\"ache}, Math. Ann. {\bf 104}, 637 (1931).
\bibitem{ryder}L.H. Ryder, {\it Dirac monopoles and the Hopf map $S^{3}$ to $S^{2}$}, J. Phys. A {\bf 13}, 437 (1980).
\bibitem{raja} R. Rajaraman, {\it Solitons and Instantons} (North-Holland, 1987).
\bibitem{proceeding} L.J. Boya, J. Mateos, {\it Group Theoretical Methods in Physics} (Springer, 1980), Editor: K.B. Wolf.
\bibitem{skyrme61b} T.H.R. Skyrme, {\it  Particle states of a quantized meson field}, Proc. R. Soc. Lond. A {\bf 262}, 237 (1961).
\bibitem{bjorken64} J.D. Bjorken, S.D. Drell, {\it Relativistic Quantum Mechanics} (McGraw-Hill, 1964).
\bibitem{hong22massive} S.T. Hong, {\it Dirac type relativistic quantum mechanics for massive photons}, Nucl. Phys. B {\bf 980}, 
115852 (2022), arXiv:2108.07299.
\end{thebibliography}
\end{document}